\documentclass[12pt]{article}
\usepackage{amsfonts}
\usepackage{bbm}
\usepackage{graphicx}
\usepackage{mathrsfs}
\usepackage{color}
\usepackage{geometry}             
\usepackage{amssymb}
\usepackage{amsmath}
\usepackage{epstopdf}
\usepackage{cases}
\usepackage{cite}

\usepackage[hyperindex=true,
          pdfstartview=FitH,
          bookmarksnumbered=true,
          bookmarksopen=true,
          citecolor=blue,
          linkcolor=blue,
          colorlinks=true,
          pdfborder=001,
          unicode]{hyperref}

\allowdisplaybreaks

\begin{document}

\title{Ricci polynomial gravity}
\author{Xin Hao
and Liu Zhao\thanks{Correspondence author.}\\
School of Physics, Nankai University, Tianjin 300071, China
\\
{\em email}: \href{mailto:shanehowe@mail.nankai.edu.cn}
{shanehowe@mail.nankai.edu.cn}
and
\href{mailto:lzhao@nankai.edu.cn}{lzhao@nankai.edu.cn}
}
\date{}
\maketitle

\begin{abstract}
We study a novel class of higher curvature gravity models in $n$ spacetime dimensions
which we call Ricci polynomial gravity. The action consists purely of a polynomial in
Ricci curvature of order $N$. In the absence of the second order terms in the action,
the models are ghost free around the Minkowski vacuum. By appropriately choosing the coupling
coefficients in front of each terms in the action, it is shown that the models can have
multiple vacua with different effective cosmological constants, and can be made free of ghost
and scalar degrees of freedom around at least one of the maximally symmetric vacua
for any $n>2$ and any $N\geq 4$. We also discuss some of the physical implications of the
existence of multiple vacua in the contexts of black hole physics and cosmology.

\end{abstract}
\newpage

\section{Introduction}

In this work, we study a particular class of higher curvature gravity models
which we call Ricci polynomial gravity. Models of higher curvature gravity have been extensively
studied for at least two reasons. One reason is to improve the ultra violet behavior
in the quest for a consistent quantum theory of gravity. The other one comes from
cosmological studies, where the higher curvature corrections play a role for the
remedy of the extra unknown source which is often called dark energy.
Of all models of higher curvature gravity, the most well studied examples are $f(R)$ gravity
\cite{Bergmann,Buchdahl,Starobinsky,Felice,Nojiri} and Lanczos-Lovelock gravity
\cite{Lanczos,Lovelock,Charmousis}. Other important models include the
so-called critical gravity \cite{Lu1,Deser0}, Conformal or Weyl gravity \cite{Weyl,O'Raifeartaigh,Fradkin}
and the various versions of massive gravities \cite{vanDam:1970vg,Zakharov:1970cc,Deser1,Deser2,Bergshoeff,Li}.
Despite the fact that many of the models behave well in one aspect or another, many of them
contain ghost or propagating massive degrees of freedom. In fact, the ghost free conditions
and the removal of massive degrees of freedom
impose very strong constraints in model construction for higher curvature gravities.
The actions of the class of models which we will
study have a very particular structure: the whole Lagrangian density consists
of a polynomial of order $N$ in Ricci curvature, with the $k$-th order term being
$\overset{(k)}{\mathcal{R}}=R^{\mu_1}{}_{\mu_2}
R^{\mu_2}{}_{\mu_3}\cdots R^{\mu_k}{}_{\mu_1}$. By properly choosing the coefficients
in front of each terms in the Lagrangian density, the models can be made free of
ghost as well as scalar degrees of freedom around at least one of the permitted vacua in
any spacetime dimension $n>2$ and for any order $N\geq 4$.  To our knowledge, this type
of higher curvature gravity has never been studied in the literature before.

\section{Field equation and vacuum solutions}

Before rushing into the action of the Ricci polynomial gravity, let us first prepare
some notations for later convenience.

For any nonnegative integer $k$, we define the tensors
$\overset{(k)}{\mathcal{R}}_{\mu\nu}$ on the spacetime manifold
$(\mathcal{M},g_{\mu\nu})$,
\begin{align*}
& \overset{(0)}{\mathcal{R}}_{\mu\nu} = g_{\mu\nu},  \\
& \overset{(1)}{\mathcal{R}}_{\mu\nu} = R_{\mu\nu},  \\
& \overset{(k)}{\mathcal{R}}_{\mu\nu}
= \overset{(k-1)}{\mathcal{R}}_{\mu\sigma} R^\sigma{}_\nu,
\end{align*}
where $R_{\mu\nu}$ is the Ricci tensor. Clearly $\overset{(k)}{\mathcal{R}}_{\mu\nu}$
is a symmetric rank 2 tensor. We define $\overset{(k)}{\mathcal{R}}$ as the trace
of $\overset{(k)}{\mathcal{R}}_{\mu\nu}$:
\[
\overset{(k)}{\mathcal{R}}
= g^{\mu\nu} \overset{(k)}{\mathcal{R}}_{\mu\nu}.
\]

Now let us consider the Ricci polynomial gravity of order $N$ in $n$ spacetime dimensions,
which may be denoted the RicPoly${}_{(n,N)}$ model for short.
The action is given as
\begin{align}
S = \frac{1}{2\kappa} \int \mathrm{d}^n x \sqrt{|g|}
\sum^N_{k=1} \alpha_k \overset{(k)}{\mathcal{R}}. \label{action}
\end{align}
There is a redundancy in parameters because $\frac{1}{2\kappa}$ is an overall 
multiplying factor which may be absorbed by rescaling all $\alpha_k$ with this factor. 
To remove this redundancy we fix $\alpha_1$ to be unity, so that 
if we choose $\{\alpha_k\}=\{1,0,\cdots,0\}$ the standard general
relativity is recovered.

The first order variation of the action reads
\begin{align}
\delta S
=& \frac{1}{2\kappa}\int \mathrm{d}^n x \sqrt{|g|}
\Big(\sum^N_{k=1} \alpha_k
\overset{(k)}{\mathcal{H}}_{\mu\nu}\Big) \delta g^{\mu\nu}
+ \frac{1}{2\kappa}\int \mathrm{d}^n x \sqrt{|g|}
\nabla^\sigma \Big(\sum^N_{k=1} k \alpha_k
\overset{(k)}{\mathcal{B}}_\sigma \Big), \label{dS}
\end{align}
where $\mathcal{B}_\sigma$ is some complicated vector expression which is linear
in $\delta g_{\mu\nu}$ and its covariant derivatives, and
\begin{align}
\overset{(k)}{\mathcal{H}}_{\mu\nu} =
k \overset{(k)}{\mathcal{R}}_{\mu\nu}
- \frac{1}{2} \overset{(k)}{\mathcal{R}} g_{\mu\nu}
+ \frac{k}{2} \big(\square \overset{(k-1)}{\mathcal{R}}_{\mu\nu}
+ g_{\mu\nu} \nabla_\alpha \nabla_\beta
\overset{(k-1)}{\mathcal{R}}{}^{\alpha\beta}\big)
- k \nabla^\sigma \nabla_{(\mu}
\overset{(k-1)}{\mathcal{R}}{}_{\nu)\sigma},  \label{Hform1}
\end{align}
which are divergence-free for all $k\geq 1$,
\begin{align}
\nabla^\mu \overset{(k)}{\mathcal{H}}_{\mu\nu} = 0. \label{divfree}
\end{align}
Assuming that we can add an appropriate boundary counter term to cancel the
extra total divergence term in \eqref{dS}, we arrive at the following
field equation,
\begin{align}
H_{\mu\nu} =
\sum^N_{k=1} \alpha_k \overset{(k)}{\mathcal{H}}_{\mu\nu} = 0.
\label{fe}
\end{align}
The divergence-free condition \eqref{divfree} ensures that we can supplement
the original action \eqref{action} by ordinary minimally coupled matter sources.

Now retaining to the sourceless field equation, it is
obvious to see that all solutions of the equation $R_{\mu\nu} = \chi g_{\mu\nu}$
are also solutions of \eqref{fe}, with the constant
$\chi$ determined by
\begin{align}
A_{(n,N)}(\chi) = 0,  \qquad
A_{(n,N)}(x)\equiv\sum^N_{k=1} \alpha_k
\big(k-\frac{n}{2}\big) x^k. \label{Aseries}
\end{align}
For arbitrarily chosen coefficients $\{\alpha_k, k\geq2\}$, nonzero solutions to \eqref{Aseries}
are not guaranteed to exist. Nevertheless, $\chi=0$ is always a solution, which means that
Ricci flat manifolds (including the Minkowski spacetime) are always vacuum solutions of
Ricci polynomial gravity. We will see in the next section that the ghost free condition
does not allow all the coefficients $\alpha_k$ to be chosen arbitrarily. On the contrary,
they must be subject to a set of algebraic constraints, which, when fulfilled, do allow
for nonzero $\chi$ to arise as solution to \eqref{Aseries}. This would, in turn, mean that
the RicPoly${}_{(n,N)}$ model may be ghost free and possess multiple vacua with at least some
vacuum having nonzero effective cosmological constant. This explains why we did not include
a bare cosmological constant term (i.e. the $k=0$ term) in the action \eqref{action}.

\section{Linear perturbations}

Assume that the field equation admits an Einstein manifold with metric $\bar g_{\mu\nu}$
satisfying
\begin{align}
R_{\mu\nu}(\bar g)=\chi \bar g_{\mu\nu} \label{bgEin}
\end{align}
as a vacuum,
and let us consider fluctuations around this background metric.

The spacetime metric with fluctuation is denoted
$g_{\mu\nu} = \bar g_{\mu\nu} + \delta g_{\mu\nu}$, where $\delta g_{\mu\nu}$ is to be
considered as ``small'' deviation from the background metric.
It is customary to denote $\delta g_{\mu\nu}$ as $h_{\mu\nu}$, and also
$h^{\mu\nu}=-\delta g^{\mu\nu} = -\bar g^{\mu\rho}\bar g^{\nu\sigma} h_{\rho\sigma}$,
$h=h^\mu{}_\mu$, $A_\mu = \nabla^\nu h_{\mu\nu}$, where $\nabla_\mu$ is the covariant
derivative associated with the background metric $\bar g_{\mu\nu}$.

Expanding the field equation to linear order in $h_{\mu\nu}$, we get
\begin{align}
\delta H_{\mu\nu}
&= \sum_{k=1}^{N} \alpha_k
\Big[ k \delta \overset{(k)}{\mathcal{R}}_{\mu\nu}
- \frac{n}{2} \chi^k h_{\mu\nu} - \frac{k}{2}
\chi^{k-1} \bar{g}_{\mu\nu}
(\nabla_\sigma A^\sigma - \square h - \chi h) \Big] \nonumber \\
&\quad+ \sum^{N-1}_{k=1} \alpha_{k+1} \Big[ \frac{k+1}{2} \big(\Box \delta \overset{(k)}{\mathcal{R}}_{\mu\nu}
+ \bar g_{\mu\nu} \nabla^\alpha \nabla^\beta
\delta \overset{(k)}{\mathcal{R}}_{\alpha\beta}\big)
- (k+1) \nabla^\sigma \nabla_{(\mu} 
\delta \overset{(k)}{\mathcal{R}}{}_{\nu)\sigma}\Big]
= 0, \label{lin1}
\end{align}
where
\begin{align}
\delta \overset{(k)}{\mathcal{R}}_{\mu\nu}
&= - \Big( \frac{1}{2} \sum^{k}_{i=1}
\overset{(i-1)}{\mathcal{R}}_\mu{}^\alpha
\overset{(k-i)}{\mathcal{R}}_\nu{}^\beta \Delta_L
+ \sum^{k-1}_{i=1} \overset{(i)}{\mathcal{R}}_\mu{}^\alpha
\overset{(k-i)}{\mathcal{R}}_\nu{}^\beta \Big) h_{\alpha\beta},
 \label{dR2} \\
\Delta_L h_{\mu\nu}
&= \square h_{\mu\nu}
+ \nabla_\mu \nabla_\nu h -2 \big(\nabla_{(\mu} A_{\nu)}
+ R^\rho{}_{(\mu}h_{\nu)\rho}
- R_\mu{}^{\alpha}{}_{\nu}{}^{\beta}
h_{\alpha\beta}\big), \label{dR1}
\end{align}
and all curvature tensors appearing in \eqref{lin1}-\eqref{dR2} are evaluated on the background
geometry. The operator $\Delta_L$ is known as the Lichnerowicz operator. Using \eqref{bgEin}, we
can simplify \eqref{dR2} into the form
\begin{align}
\delta \overset{(k)}{\mathcal{R}}_{\mu\nu}
= - \Big(\frac{1}{2} k \chi^{k-1} \Delta_L
+ (k-1) \chi^k \Big) h_{\mu\nu}.  \label{dREin}
\end{align}
Further, if we take the background spacetime to be a maximally symmetric manifold with
\begin{align}
R_{\mu\alpha\nu\beta}(\bar g) = \frac{\chi}{n-1}
(\bar g_{\mu\nu}\bar g_{\alpha\beta}-\bar g_{\mu\beta}\bar g_{\alpha\nu}),
\quad R_{\mu\nu}(\bar g) = \chi \bar g_{\mu\nu},
\quad R(\bar g) = n \chi, \label{maxsym}
\end{align}
eq.\eqref{dR1} can be simplified into
\begin{align}
& \Delta_L h_{\mu\nu}
= \square h_{\mu\nu} + \nabla_\mu \nabla_\nu h
- 2 \nabla_{(\mu} A_{\nu)} -  \frac{2n\chi}{n-1}
h_{\mu\nu} + \frac{2\chi}{n-1} h \bar g_{\mu\nu}. \label{Lichmax}
\end{align}
Inserting \eqref{Lichmax} into \eqref{dREin} and then into \eqref{lin1}, we will get the fully
expanded linear perturbation equation.

We need to proceed differently for background vacua with $\chi=0$ and $\chi\neq 0$.

\textbf{(a) Minkowski background}

If $\chi = 0$, i.e. Minkowski background, eq.\eqref{lin1} will be significantly
simplified, yielding
\begin{align}
& \alpha_2 \big[\square^2 h_{\mu\nu}
+ \eta_{\mu\nu} \square (\square h
- \partial^\sigma A_\sigma)
- \square \partial_\mu \partial_\nu h
-2 \square \partial_{(\mu} A_{\nu)}
+2 \partial_\mu\partial_\nu \partial^\sigma A_\sigma\big] \nonumber\\
&\qquad +\big[\square h_{\mu\nu} - \eta_{\mu\nu}
(\square h - \partial^\sigma A_\sigma)
+ \partial_\mu \partial_\nu h
-2 \partial_{(\mu} A_{\nu)}\big] = 0,  \label{le}
\end{align}
where the second line corresponds to the $\alpha_1(=1)$ terms.
The absence of $\alpha_k (k>2)$ in \eqref{le} implies that
the terms of order $k>2$ do not affect the perturbations around the Minkowski background,
the only complexity comes from the quadratic Ricci tensor term.

If, in addition, we let $\alpha_2 = 0$, the linearized field equation \eqref{le} will be
the same as the linearized standard Einstein equation, which describes a normal massless
spin-2 field and so the fluctuation around the Minkowski vacuum is ghost free.

If $\alpha_2 \neq 0$, and for example when $n = 4$, we can
decompose $h_{\mu\nu}$ into its trace $h$ and traceless parts by defining
\[
\bar{h}_{\mu\nu} = h_{\mu\nu}
- \frac{1}{4} h \eta_{\mu\nu},
\]
and choose the gauge
\[
\partial^\nu \bar{h}_{\mu\nu}
= A_\mu - \frac{1}{4} \partial_\mu h = 0,
\]
then the linearized field equation \eqref{le} becomes
\begin{align}
\square \Big(\square + \frac{1}{\alpha_2}\Big)
\bar{h}_{\mu\nu} + (\eta_{\mu\nu} \square
- \partial_\mu\partial_\nu)
\Big(\square - \frac{1}{2 \alpha_2}\Big) h =0,
\end{align}
from which we see that either the trace part or the traceless part contains a ghost degree of
freedom. The same is true had we chosen any other spacetime dimensions. Combined together,
we conclude that $\alpha_2 = 0$ should be the ghost free condition in the case
with $\chi = 0 $.

\textbf{(b) (A)dS background}

If $\chi \neq 0$, it will be convenient to work in the gauge $A_\mu = \nabla_\mu h$.
The trace of eq.\eqref{lin1} gives
\begin{align}
\big[ B_{(n,N)}(\chi) \square
+ \chi A'_{(n,N)}(\chi) \big] h =0,  \label{tr}
\end{align}
where $A'_{(n,N)}(\chi)$ is the derivative of $A_{(n,N)}(x)$ with respect to $x$
evaluated at $x=\chi$, and
\[
B_{(n,N)}(\chi)
=\frac{1}{4}\sum^{N-1}_{k=1} \alpha_{k+1}(k+1) \big[n(k-2)+2\big] \chi^k.
\]
So, as long as $A'_{(n,N)}(\chi) \neq 0$, eq. \eqref{tr} will describe a
propagating scalar mode except when
\[
B_{(n,N)}(\chi) = 0,
\]
in which case we learn from eq \eqref{tr} that $h =0$, and thus
$A_\mu = \nabla^\nu h_{\mu\nu} = 0$, i.e. $h_{\mu\nu}$ is
traceless and transverse, therefore eq. \eqref{lin1} becomes
\begin{align}
\delta H_{\mu\nu}& =-\Big\{\frac{C_{(n,N)}(\chi)}{4\chi} \Box^2
+ \Big[\frac{n-1}{2} - C_{(n,N)}(\chi)\Big]
\frac{1}{n-1} \Box    \nonumber \\
&\qquad \quad
+ \Big[\big(1-\frac{n}{2}\big)
C_{(n,N)}(\chi) - (n-1) \Big]
\frac{\chi}{(n-1)^2} \Big\} h_{\mu\nu}= 0,
 \label{h1}
\end{align}
where
\[
C_{(n,N)}(\chi) = \sum^{N-1}_{k=1} \alpha_{k+1}
k (k+1) \chi^k.
\]
If $C_{(n,N)}(\chi) = 0$, we get
\begin{align}
\Big(\Box - \frac{2\chi}{n-1} \Big) h_{\mu\nu} =0, \label{h2}
\end{align}
which describes a massless spin-2 mode; if
$C_{(n,N)}(\chi) \neq 0$, eq. \eqref{h1} can be rearranged in the form
\begin{align}
\Big(\Box - \frac{2\chi}{n-1} - m_1^2\Big)
\Big(\Box - \frac{2\chi}{n-1} - m_2^2\Big)
h_{\mu\nu} =0, \label{h3}
\end{align}
with the two mass parameters $m_{1,2}$ satisfying
\[
m_1^2 m_2^2=-\frac{n}{2} \Big(\frac{2 \chi}{n-1}\Big)^2.
\]
Thus, in the case when $C_{(n,N)}(\chi) \neq 0$, we will have
a massive spin-2 field and a spin-2 ghost.

To summarize, in the case $\chi \neq 0$ and if we wish to remove the propagating
scalar mode altogether, the ghost free condition will read
\begin{align}
A_{(n,N)}(\chi)= 0, \quad
B_{(n,N)}(\chi)=0, \quad
C_{(n,N)}(\chi)= 0, \quad
A'_{(n,N)}(\chi) \neq 0. \label{GF}
\end{align}
Looking into the detailed structure of the polynomials $A_{(n,N)}(\chi), B_{(n,N)}(\chi)$
and $C_{(n,N)}(\chi)$, we will see that whenever the first 3 equalities in \eqref{GF}
are satisfied, we will have
\[
A'_{(n,N)}(\chi)=1-\frac{n}{2} < 0 \qquad(\mbox{for }n>2).
\]
So the last condition is not really necessary.

An equivalent form of the ghost free condition
\eqref{GF} can be written as
\begin{align}
\sum^{N-1}_{k=1} \alpha_{k+1} \chi^k
= -\sum^{N-1}_{k=1} \alpha_{k+1} k \chi^k
= \sum^{N-1}_{k=1} \alpha_{k+1} k^2 \chi^k
= \frac{2}{n}-1.   \label{GF2}
\end{align}
For any $N>1$, this amounts to solve $N-1$ coefficients out of 3 equations. Therefore, for
$N<4$, \eqref{GF2} will give rise to a system of over determined algebraic equations for the
coefficients $\{\alpha_k, 2\leq k \leq N\}$, and it turns out that such a system of over
determined equations is inconsistent. If $N=4$, \eqref{GF2} constitutes a system of determined
algebraic equations, whose solution is characterized purely by the value of $n$ and $\chi$.
If $N>4$, \eqref{GF2} constitutes a system of undetermined equations, which implies that
$N-4$ of the coefficients $\{\alpha_k\}$ can be chosen arbitrarily. Since $\chi$ itself is
to be considered as a free parameter in the action, there will be as many as $N-3$ free
parameters in the ghost free RicPoly${}_{(n,N)}$ models.

It is worth noting that the above ghost free conditions ensure only the removal of ghost
fluctuations around the vacuum \eqref{maxsym}. However, since $A_{(n,N)}(x)$ is a polynomial of
order $N$ in $x$, it can have up to $N$ roots in principle, and each of the roots gives a 
vacuum for the RicPoly${}_{(n,N)}$ model. So, the removal of ghost fluctuations
around one of the vacua does not imply the removal of ghost fluctuations around other vacua.
For $N>4$ models we can take the liberty of deliberately choosing the values of the extra
$N-4$ free parameters to make the model ghost free also at some other vacua. For instance,
if $N=5$ we can set $\alpha_2=0$ to make the model also ghost free around the Minkowski vacuum.
Following the same spirits, for models of even higher order $N$ in Ricci curvature, 
we can take the liberty of choosing the free parameters $\alpha_2, \cdots, \alpha_{N-3}$ 
to make the model ghost free around as many as possible vacua, however, 
it is evident that it is impossible to make the model ghost free around all vacua.
In any case, we think of the fact that the RicPoly${}_{(n,N)}$ model can be ghost free around
some of the vacua but not around all of them as a feature rather than defect, because,
in cosmology, ghost like degrees of freedom are sometimes necessary to facilitate
accelerating expansion.

Tables \ref{tab1} through \ref{tab3} give the list of coupling coefficients for all
the RicPoly${}_{(n,N)}$ models of order $N=4,5,6$ in dimensions $n=3,4,\cdots,10$
which are ghost and scalar free around the maximally symmetric vacuum described by
\eqref{maxsym} with $\chi\neq 0$. We did not fix the free parameter $\alpha_2$ in the
case $N=5$ and $\alpha_2,\alpha_3$ in the case $N=6$, because there is no obvious reasons
to make the model ghost free around which extra vacua besides the maximally symmetric one
given by \eqref{maxsym}.

\begin{table}[htbp]
\begin{center}\begin{tabular}{cccc}
\hline
\hline
$n$  & $\alpha_2         $ & $\alpha_3           $ & $\alpha_4           $ \\
\hline
  3  & $-\frac{2}{\chi}  $ & $\frac{8}{3\chi^2}  $ & $-\frac{1}{\chi^3}  $ \\
  4  & $-\frac{3}{\chi}  $ & $\frac{4}{\chi^2}   $ & $-\frac{3}{2\chi^3} $ \\
  5  & $-\frac{18}{5\chi}$ & $\frac{24}{5\chi^2} $ & $-\frac{9}{5\chi^3} $ \\
  6  & $-\frac{4}{\chi}  $ & $\frac{16}{3\chi^2} $ & $-\frac{2}{\chi^3}  $ \\
  7  & $-\frac{30}{7\chi}$ & $ \frac{40}{7\chi^2}$ & $-\frac{15}{7\chi^3}$ \\
  8  & $-\frac{9}{2\chi} $ & $ \frac{6}{\chi^2}  $ & $-\frac{9}{4\chi^3} $ \\
  9  & $-\frac{14}{3\chi}$ & $ \frac{56}{9\chi^2}$ & $-\frac{7}{3\chi^3} $ \\
  10 & $-\frac{24}{5\chi}$ & $\frac{32}{5\chi^2} $ & $-\frac{12}{5\chi^3}$ \\
\hline
\hline
\end{tabular} \caption{Coupling coefficients for $N=4$ models}\label{tab1}
\end{center}
\end{table}

\begin{table}[htbp]
\begin{center}\begin{tabular}{cccc}
\hline
\hline
$n$  & $\alpha_3             $ & $\alpha_4             $ & $\alpha_5           $ \\
\hline
  3  & $-\frac{3\alpha_2}{\chi}-\frac{10}{3\chi^2}   $ &
       $\frac{3\alpha_2}{\chi^2}+\frac{5}{\chi^3}    $ &
       $-\frac{\alpha_2}{\chi^3}-\frac{2}{\chi^4}    $ \\
  4  & $-\frac{3\alpha_2}{\chi}-\frac{5}{\chi^2}     $ &
       $\frac{3\alpha_2}{\chi^2}+\frac{15}{2\chi^3}  $ &
       $-\frac{\alpha_2}{\chi^3}-\frac{3}{\chi^4}    $ \\
  5  & $-\frac{3\alpha_2}{\chi}-\frac{6}{\chi^2}     $ &
       $\frac{3\alpha_2}{\chi^2}+\frac{9}{\chi^3}    $ &
       $-\frac{\alpha_2}{\chi^3}-\frac{18}{5\chi^4}  $ \\
  6  & $-\frac{3\alpha_2}{\chi}-\frac{20}{3\chi^2}   $ &
       $\frac{3\alpha_2}{\chi^2}+\frac{10}{\chi^3}   $ &
       $-\frac{\alpha_2}{\chi^3}-\frac{4}{\chi^4}    $ \\
  7  & $-\frac{3\alpha_2}{\chi}-\frac{50}{7\chi^2}   $ &
       $\frac{3\alpha_2}{\chi^2}+\frac{75}{7\chi^3}  $ &
       $-\frac{\alpha_2}{\chi^3}-\frac{30}{7\chi^4}  $ \\
  8  & $-\frac{3\alpha_2}{\chi}-\frac{15}{2\chi^2}   $ &
       $\frac{3\alpha_2}{\chi^2}+\frac{45}{4\chi^3}  $ &
       $-\frac{\alpha_2}{\chi^3}-\frac{9}{2\chi^4}   $ \\
  9  & $-\frac{3\alpha_2}{\chi}-\frac{70}{9\chi^2}   $ &
       $\frac{3\alpha_2}{\chi^2}+\frac{35}{3\chi^3}  $ &
       $-\frac{\alpha_2}{\chi^3}-\frac{14}{3\chi^4}  $ \\
  10 & $-\frac{3\alpha_2}{\chi}-\frac{8}{\chi^2}     $ &
       $\frac{3\alpha_2}{\chi^2}+\frac{12}{\chi^3}   $ &
       $-\frac{\alpha_2}{\chi^3}-\frac{24}{5\chi^4}  $ \\
\hline
\hline
\end{tabular} \caption{Coupling coefficients for $N=5$ models}\label{tab2}
\end{center}
\end{table}

\begin{table}[htbp]
\begin{center}\begin{tabular}{cccc}
\hline
\hline
$n$  & $\alpha_4                                   $ &
       $\alpha_5                                   $ &
       $\alpha_6                                   $ \\
\hline
  3  & $-\frac{6\alpha_2}{\chi^2}-\frac{3\alpha_3}{\chi}-\frac{5}{\chi^3}   $ &
       $\frac{8\alpha_2}{\chi^3}+\frac{3\alpha_3}{\chi^2}+\frac{8}{\chi^4}  $ &
       $-\frac{3\alpha_2}{\chi^4}-\frac{\alpha_3}{\chi^3}-\frac{10}{3\chi^5}$ \\
  4  & $-\frac{6\alpha_2}{\chi^2}-\frac{3\alpha_3}{\chi}-\frac{15}{2\chi^3} $ &
       $\frac{8\alpha_2}{\chi^3}+\frac{3\alpha_3}{\chi^2}+\frac{12}{\chi^4} $ &
       $-\frac{3\alpha_2}{\chi^4}-\frac{\alpha_3}{\chi^3}-\frac{5}{\chi^5}  $ \\
  5  & $-\frac{6\alpha_2}{\chi^2}-\frac{3\alpha_3}{\chi}-\frac{9}{\chi^3}   $ &
       $\frac{8\alpha_2}{\chi^3}+\frac{3\alpha_3}{\chi^2}+\frac{72}{5\chi^4}$ &
       $-\frac{3\alpha_2}{\chi^4}-\frac{\alpha_3}{\chi^3}-\frac{6}{\chi^5}  $ \\
  6  & $-\frac{6\alpha_2}{\chi^2}-\frac{3\alpha_3}{\chi}-\frac{10}{\chi^3}  $ &
       $\frac{8\alpha_2}{\chi^3}+\frac{3\alpha_3}{\chi^2}+\frac{16}{\chi^4} $ &
       $-\frac{3\alpha_2}{\chi^4}-\frac{\alpha_3}{\chi^3}-\frac{20}{3\chi^5}$ \\
  7  & $-\frac{6\alpha_2}{\chi^2}-\frac{3\alpha_3}{\chi}-\frac{75}{7\chi^3} $ &
       $\frac{8\alpha_2}{\chi^3}+\frac{3\alpha_3}{\chi^2}+\frac{120}{7\chi^4}$ &
       $-\frac{3\alpha_2}{\chi^4}-\frac{\alpha_3}{\chi^3}-\frac{50}{7\chi^5}$ \\
  8  & $-\frac{6\alpha_2}{\chi^2}-\frac{3\alpha_3}{\chi}-\frac{45}{4\chi^3} $ &
       $\frac{8\alpha_2}{\chi^3}+\frac{3\alpha_3}{\chi^2}+\frac{18}{\chi^4} $ &
       $-\frac{3\alpha_2}{\chi^4}-\frac{\alpha_3}{\chi^3}-\frac{15}{2\chi^5}$ \\
  9  & $-\frac{6\alpha_2}{\chi^2}-\frac{3\alpha_3}{\chi}-\frac{35}{3\chi^3} $ &
       $\frac{8\alpha_2}{\chi^3}+\frac{3\alpha_3}{\chi^2}+\frac{56}{3\chi^4}$ &
       $-\frac{3\alpha_2}{\chi^4}-\frac{\alpha_3}{\chi^3}-\frac{70}{9\chi^5}$ \\
  10 & $-\frac{6\alpha_2}{\chi^2}-\frac{3\alpha_3}{\chi}-\frac{12}{\chi^3}  $ &
       $\frac{8\alpha_2}{\chi^3}+\frac{3\alpha_3}{\chi^2}+\frac{96}{5\chi^4}$ &
       $-\frac{3\alpha_2}{\chi^4}-\frac{\alpha_3}{\chi^3}-\frac{8}{\chi^5}  $ \\
\hline
\hline
\end{tabular} \caption{Coupling coefficients for $N=6$ models} \label{tab3}
\end{center}
\end{table}

It is illustrative to consider the energy for the massless tensor perturbation mode \eqref{h2},
which may be evaluated via calculating the second order variation of the action \eqref{action}
under the maximally symmetric background \eqref{maxsym}. To be more concrete, since the first
variation of the action \eqref{action} reads
\[
\delta S = \frac{1}{2 \kappa}
\int \mathrm{d}^n x \sqrt{|g|}
H_{\mu\nu} \delta g^{\mu\nu}
= -\frac{1}{2 \kappa}
\int \mathrm{d}^n x \sqrt{|g|}
H_{\mu\nu} h^{\mu\nu},
\]
the second variation can be written as
\[
\delta^2 S
= -\frac{1}{2 \kappa}
\int \mathrm{d}^n x \sqrt{|g|}\, \Big(h^{\mu\nu} \delta H_{\mu\nu}+\delta h^{\mu\nu} H_{\mu\nu}
+\frac{1}{2}H_{\alpha\beta} h^{\alpha\beta}g^{\mu\nu}h_{\mu\nu}
\Big),
\]
where the last two terms can be dropped because $H_{\mu\nu}$ vanishes due to the field equation,
and $\delta H_{\mu\nu}$ in the first term is given by \eqref{h1}, if the background metric is
chosen to be \eqref{maxsym}. Under our choice of ghost free
condition $C_{(n,N)}(\chi) = 0$, we have
\[
\delta H_{\mu\nu}
= - \frac{1}{2}\Big(
\Box - \frac{2 \chi}{n-1} \Big) h_{\mu\nu},
\]
and hence,
\begin{align*}
\delta^2 S
&= -\frac{1}{2 \kappa}
\int \mathrm{d}^n x \sqrt{|\bar{g}|}
(h^{\mu\nu} \delta H_{\mu\nu})  \\
&= -\frac{1}{2 \kappa}
\int \mathrm{d}^n x \sqrt{|\bar{g}|}
\Big( \frac{1}{2} \nabla^{\sigma} h^{\mu\nu}
\nabla_\sigma h_{\mu\nu}
+ \frac{\chi}{n-1} h^{\mu\nu} h_{\mu\nu} \Big)
\equiv \int \mathrm{d}^n x \mathcal{L}_2,
\end{align*}
where the second equality holds up to a total divergence.

The momentum density that is conjugates to $h_{\mu\nu}$ reads
\begin{align*}
\pi^{\mu\nu} =
\frac{\delta \mathcal{L}_2}{\delta(\dot{h}_{\mu\nu})}
= - \frac{1}{2 \kappa} \sqrt{|\bar{g}|}
\nabla^0 h^{\mu\nu}.
\end{align*}
Therefore the canonical Hamiltonian is
\begin{align*}
H &= \int \mathrm{d}^{n-1} x (\pi^{\mu\nu} \dot{h}_{\mu\nu}
- \mathcal{L}_2)
= - \frac{1}{2\kappa} \int \mathrm{d}^{n-1} \sqrt{|\bar g|}
(\nabla^0 h^{\mu\nu} \dot{h}_{\mu\nu}
- h^{\mu\nu} \delta H_{\mu\nu})\\
&= - \frac{1}{2\kappa} \int \mathrm{d}^{n-1} \sqrt{|\bar g|}
\nabla^0 h^{\mu\nu} \dot{h}_{\mu\nu},
\end{align*}
where in the last step we have used the perturbation equation to set $\delta H_{\mu\nu}=0$.
The final result for the Hamiltonian is then identified as the energy of the perturbation mode,
which is the same as that for standard Einstein gravity and was known
to be positive \cite{Lu1}.

It remains to consider the case $B_{(n,N)}(\chi)\neq 0$, i.e. when the propagating scalar mode
is present. In this case, eq.\eqref{tr} can be rearranged in the form
\[
\left(\square+\frac{\chi A'_{(n,N)}(\chi)}{B_{(n,N)}(\chi)}\right)h=0,
\]
therefore, depending on the value of the expression $\frac{\chi A'_{(n,N)}(\chi)}
{B_{(n,N)}(\chi)}$, $h$ may corresponds to either a massive/massless scalar mode, or a scalar ghost 
mode. Identifying the complete ghost free conditions when the
scalar mode is present will be a complicated task and we shall postpone it to future works.

\section{RicPoly${}_{(4,4)}$: black hole and cosmological solutions}

Among the various choices for $N$ and $n$, the simplest and most promising example case
which is ghost free at least around one of the vacua with $\chi\neq 0$ is RicPoly${}_{(4,4)}$.
The action for this simplest model reads
\begin{align}
S = \frac{1}{2\kappa}\int \mathrm{d}^4 x \sqrt{|g|}
\Big(R
- \frac{3}{\chi}\overset{(2)}{\mathcal{R}}
+ \frac{4}{\chi^2}\overset{(3)}{\mathcal{R}}
-\frac{3}{2\chi^3} \overset{(4)}{\mathcal{R}} \Big). \label{poly44}
\end{align}

The vacuum structure of this model is governed by the zeros of the polynomial
$A_{(4,4)}(x)$, which, under the ghost free condition, reads
\[
A_{(4,4)}(x)=-\frac{3x^4}{\chi^3}+\frac{4x^3}{\chi^2}-x.
\]
In this case, there are 4 real roots
\begin{align}
x_i\equiv \Lambda_i=0,\chi,\frac{1}{6}(1+\sqrt{13})\chi,\frac{1}{6}(1-\sqrt{13})\chi, \label{cosmo}
\end{align}
each corresponds to an Einstein manifold $R_{\mu\nu}^{(i)}=\Lambda_i g_{\mu\nu}^{(i)}$
($i=i,2,3,4$) as a vacuum solution for the model \eqref{poly44}. Note that these are not
necessarily the same as the maximally symmetric vacua around which we made linear perturbations
in the last section. Instead, they can be generic Einstein manifolds with effective
cosmological constants $\Lambda_i$. By the way, let us mention that the existence of multiple vacua 
has also been observed in \cite{Boulware:1985wk}, which, of course studies a different model 
of extended gravity.  

To be more specific, let us consider static spherically symmetric black hole solutions. The
usual Schwarzschild-(A)dS black holes fall into this class of solutions:
\begin{align}
&\mathrm{d}s^2=-f_i(r)\mathrm{d}t^2
+f_i(r)^{-1}\mathrm{d}r^2
+r^2\mathrm{d}\Omega_2^2,\\
&f_i(r)=1-\frac{2M}{r}-\frac{\Lambda_i r^2}{3}.
\end{align}
This looks all the same like the standard general relativity with a cosmological constant.
What makes a big difference is that there are simultaneously 4 such solutions with the same mass
parameter $M$ but different $\Lambda_i$ given in \eqref{cosmo}. Since the radii of the event
horizons of the black holes are determined by the equations
\[
f_i(r)=0,
\]
it is clear that black holes of the same mass in Ricci polynomial gravity can have different
radii. Consequently, when considering the thermodynamic behaviors of these black holes,
the temperatures should be different for different solutions. Moreover, since Ricci
polynomial gravity belongs to the class of higher curvature gravities, the entropies of the
black holes should be evaluated by Wald entropy formula rather than Bekenstein-Hawking
formula, and it is naturally expected that they differ from the entropies of the black holes
with the same metric in Einstein gravity. The entropies for black holes with different
$\Lambda_i$ should also differ from each other even though the mass parameters are identical.
These features suggest that there may be rich structures in the thermodynamics - especially
concerning phase transitions - for the black hole solutions.

Before ending, let us pay a few words on the cosmological
implications of the effective cosmological constants.
It is evident from \eqref{cosmo} that provided $\chi\neq0$ (which is implied when we write the
action \eqref{poly44}), the 4 effective cosmological constants will have different signs.
For instance, if $\chi>0$, then two of the effective cosmological constants will be possitive,
one is negative and one is zero. Different signatures of $\Lambda_i$ signifies different
asymptotics of the solutions, and this can have profound implications when cosmological
solutions are being considered. For simplicity, let us present the spatially flat FRW
cosmological solutions of the field equations. The FRW metrics are given by
\begin{align}
&\mathrm{d}s^2=-\mathrm{d}t^2
+a_i(t)(\mathrm{d}x^2+\mathrm{d}y^2+\mathrm{d}z^2),\\
&a_i(t)=1,\exp\Big(\frac{2\sqrt{3\chi}}{3}t\Big),
\exp\Big(\frac{\sqrt{2\chi}}{3}\sqrt{1+\sqrt{13}}t\Big),
\exp\Big(\frac{\sqrt{2\chi}}{3}\sqrt{1-\sqrt{13}}t\Big).
\end{align}
Some of these solutions are complex and should be neglected. For instance, let us assume
$\chi>0$. Then the last solution is complex, and we are left with two
accelerating expanding solutions and one non-expanding solution. It is tempting to
consider such a possibility that the universe begins from one accelerating expanding phase,
then, after a short period, experience a phase transition to the non-expanding phase for
another period of time, and finally make a second transition to the other accelerating
phase at late time. This rough picture feels close to the current observational universe,
even without considering the contribution of ordinary matter sources. So, if refined
with ordinary matter contribution, this may led to a useful model for the
evolution of the universe.

\section{Concluding Remarks}

Our preliminary study has revealed several interesting features of the Ricci polynomial
gravities. First of all, the inclusion of terms of higher order in Ricci curvature
improves the ultra violet behavior. By power counting, such models may be super renormalizable.
Secondly, unlike other higher curvature gravity models which often develop ghost degrees of
freedom, the Ricci polynomial gravity models can be made free of ghost as well as massive
degrees of freedom at least around one of their maximally symmetric vacua, and provided the
order $N$ is high enough, the ghost free conditions are not too restrictive so that there can
be as many as $N-3$ free parameters in the action. The existence of
multiple Einstein vacua with different effective cosmological constants is the third
feature which may have profound implications in black hole physics as well as in cosmological
contexts. We hope to make deeper understandings about these models in future works.

\section*{Notes Added}

Recently, long after the first version of the present work appeared on arXiv, 
Y.~Z.~Li, H.~S.~Liu and H.~L\"u published the paper
\cite{Li:2017ncu} on arXiv, which also studies some Ricci curvature-extended gravity models.
Their models are quite different from our's, because they also included Ricci scalar factors 
in the Lagrangian. Even though, it is interesting to observe that the so-called 
quasi-topological conditions in \cite{Li:2017ncu} is actually a synonym to our
ghost free condition. In this sense, our models under the ghost free conditions actually 
belong to a simpler class of quasi-topological Ricci polynomial gravity models as compared to 
those in \cite{Li:2017ncu}.

\section*{Acknowledgement}

This work is supported by the National Natural Science Foundation of China under the grant
No. 11575088.

\providecommand{\href}[2]{#2}\begingroup
\footnotesize\itemsep=0pt
\providecommand{\eprint}[2][]{\href{http://arxiv.org/abs/#2}{arXiv:#2}}


\end{document}